\documentclass[prd,showpacs,amsmath,amssymb,superscriptaddress,nofootinbib,showkeys]{revtex4}
\usepackage{amssymb}
\usepackage{amsmath}
\usepackage[cp1251]{inputenc}
\usepackage[T2A]{fontenc}
\usepackage{graphicx}
\usepackage{floatflt}
\DeclareGraphicsExtensions{.pdf,.eps,.png,.jpg}

\begin{document}
\title{Reconstruction in the Horndeski theory within the scope of the  Bianchi I cosmology}

\author{Ruslan K. Muharlyamov}
\email{rmukhar@mail.ru} \affiliation{Department of General
Relativity and Gravitation, Institute of Physics, Kazan Federal
University, Kremlevskaya str. 18, Kazan 420008, Russia}

\author{Tatiana N. Pankratyeva}
\email{ghjkl.15@list.ru} \affiliation{Department of Higher
Mathematics, Kazan State Power Engineering University,
Krasnoselskaya str. 51, Kazan 420066, Russia}

%\date{\today}
%\preprint{\bf version 1.6}

\begin{abstract}

In the previous article Eur. Phys. J. Plus 136, 590 (2021) (arXiv:
2110.15396) we have proposed a reconstruction method for the
kinetic gravity braiding theory  in the framework of the flat
Friedman-Robertson-Walker spacetime. Here we develop this method
in the Bianchi I spacetime model for a subclass of the Horndeski
theory: $G_5 \sim \phi $, $G_2(X)\neq0$, $G_3(X)\neq0$. The Hubble
parameter $H(t)$ and the canonical kinetic term $X(t)$ are set a
priori. The choice of the function $X(t)$ determines the
anisotropic properties of the Universe. This makes it possible to
provide believable anisotropy. The presented method allows for a
realistic model of the Universe to simply reconstruct some scalar
field theory. Reconstruction example is given for anisotropic
model of a post-inflationary transition to the radiation-dominated
phase. The model is investigated for the absence ghosts and
Laplacian instabilities.

\end{abstract}



\maketitle

\section{Introduction}

To explain accelerated expansion of the Universe and other
observational facts, modifications of the gravity theory are used.
The Horndeski gravity (HG)  has interesting properties
\cite{Horndeski}. The theory is constructed in such a way that the
motion equations  are of the order of the derivative no higher
than the second. In this sense, the HG is the most general variant
of the scalar-tensor theory of gravitation. As we will see below,
the action density  has a cumbersome form and contains four
arbitrary functions. However, this provides a wide field for
action. Within the framework of the HG, various cosmological and
astrophysical problems are solved \cite{Sushkov0, Appleby0,
Sotiriou, Babichev, Maselli, Muharlyamov}.

Authors \cite{Kobayashi1} proposed such a parametrization of the
action density for the HG:
\begin{eqnarray}\label{action} L_H=\sqrt{-g}\Big(\mathcal{L}_2+\mathcal{L}_3+\mathcal{L}_4+\mathcal{L}_5\Big) \,,\end{eqnarray}
$$\mathcal{L}_2 = G_2(\phi,X)\,,\, \mathcal{L}_3 = -
G_3(\phi,X)\Box\phi\,,$$
$$\mathcal{L}_4 = G_{4}(\phi,X) R +G_{4X}(\phi,X) \left[ (\square
\phi )^{2}-(\nabla_\mu \nabla_\nu \phi)^2  \right]\,,$$
\begin{equation} \mathcal{L}_5 = G_{5}(\phi,X) G_{\mu\nu}\,\nabla^\mu \nabla^\nu
\phi -\frac{1}{6}G_{5X}   \left[\left( \Box \phi \right)^3 -3 \Box
\phi (\nabla_\mu \nabla_\nu \phi)^2 + 2\left(\nabla_\mu \nabla_\nu
\phi \right)^3 \right], \label{lagr2}
\end{equation}
where $g$ is the determinant of metric tensor $g_{\mu\nu}$; $R$ is
the Ricci scalar and $G_{\mu\nu}$ is the Einstein tensor; the
functions $G_{i}$ ($i=2,3,4,5$) depend on the scalar field $\phi$
and the canonical kinetic term, $X=-\frac{1}{2}\nabla^\mu\phi
\nabla_\mu\phi$. The following designations are used $G_{iX}\equiv
\partial G_i/\partial X$,
 $(\nabla_\mu \nabla_\nu \phi)^2\equiv\nabla_\mu \nabla_\nu \phi
\,\nabla^\nu \nabla^\mu \phi$, and $\left(\nabla_\mu \nabla_\nu
\phi \right)^3\equiv \nabla_\mu \nabla_\nu \phi  \,\nabla^\nu
\nabla^\rho \phi \, \nabla_\rho \phi  \nabla^\mu \phi$.

We continue the research begun in the work \cite{Muharlyamov},
which  considered the kinetic gravity braiding (KGB) theory
($G_2(X)\neq0, \, G_3(X)\neq0,\, G_4=1/(16\pi)$) in the flat
Friedman-Robertson-Walker spacetime (FRW). Here we present a
reconstruction method in the Bianchi I spacetime model (BI) and in
the FRW spacetime for the subclass of HG:
\begin{equation} \label{lagr3} G_2(X)\neq0, \, G_3(X)\neq0,\, G_4=1/(16\pi),
\, G_5 = \eta\cdot\phi/2 , \, \eta=\text{const}.\end{equation} The
functions $G_2$, $G_3$ are reconstructed based on the given the
form of the Hubble parameter $H(t)$ and kinetic density $X(t)$.
The choice freedom of functions $H$, $X$ allows you to regulate
the issues of model stability \cite{DeFelice1,Appleby1}. As shown
in refs. \cite{Tahara,Sushkov,SushkovStar}, the presence of the
function $G_5$ in the action density can lead to non-standard
behavior of the anisotropic Universe. During the reconstruction,
the function $X$ directly models the the anisotropy behavior   of
the Universe. This moment is important to ensure believable
anisotropy. For example, authors \cite{Hawking}, \cite{Hu} claim
that the isotropization of the Universe occurred quite early. At
the  start of primordial  nucleosynthesis, the universe must
already be isotropic.

The presented method allows for a realistic model of the Universe
to simply reconstruct some scalar field theory. For example, we
will reconstruct the field theory for a post-inflationary
transition to the radiation-dominated phase. Authors R. C.
Bernardo and I. Vega \cite{Bernardo} proposed their own version of
reconstruction the KGB theory for the FRW spacetime. In this
theory, the scalar charge is zero. To provide a dynamic solution
$H(t)$, $X(t)$, a another matter with a density $\rho(t)$ was
introduced. In our method, the only source is the scalar field
$\phi$. The dynamics of functions $H(t)$ and $X(t)$ is provided by
the nonzero scalar charge. For the BI model, there is an
additional contribution to the dynamics of $H(t)$, $X(t)$ from the
nonzero anisotropic charges. R. C. Bernardo et al \cite{Bernardo1,
Bernardo2} significantly develop the ideas of reconstruction in
connection with observational data. Authors \cite{Bernardo1}
reconstruct the parameters $H(t)$ from cosmic chronometers,
supernovae, and baryon acoustic oscillations compiled data sets
via the Gaussian process method and use it to draw out HG that are
fully anchored on expansion history data. In the work
\cite{Bernardo2}, dark energy is studied through the viewpoints of
parametric and nonparametric analyses of late-time cosmological
data.

\section{Field equations}

The metric of a Bianchi-I geometry may be written  as
\begin{equation}
 ds^2 = -dt^2 + a^2_1(t)dx^2 + a^2_2(t)dy^2 + a^2_3(t)dz^2. \label{met0}
\end{equation}
We consider a homogeneous model: $a_i=a_i(t)$ and $\phi=\phi(t)$ .
For the action  density (\ref{action}) and the Bianchi-I geometry
(\ref{met0}) the gravitation  equations have the form
\cite{SushkovStar}:
 $$G^0_0\left({\cal G}-2G_{4X}\dot{\phi}^2 -
2G_{4XX}\dot{\phi}^4 +2G_{5\phi}\dot{\phi}^2
+G_{5X\phi}\dot{\phi}^4\right) = G_2 - G_{2X}\dot{\phi}^2-
$$\begin{eqnarray} \label{00}- 3G_{3X}H\dot{\phi}^3 +
G_{3\phi}\dot{\phi}^2
  + 6G_{4\phi}H\dot{\phi} + 6G_{4X\phi}\dot{\phi}^3 H
 - 5G_{5X}H_1H_2H_3\dot{\phi}^3 - G_{5XX}H_1H_2H_3\dot{\phi}^5\,,\end{eqnarray}
$$ {\cal G}G^{i}_{i}-(H_{j}+H_{k})\frac{d{\cal G}}{dt} = G_2 - \dot\phi
\frac{d G_3}{dt} +2 \frac{d}{dt}(G_{4\phi}\dot{\phi})
-$$\begin{eqnarray} \label{ii}-\frac{d}{dt}(G_{5X}\dot{\phi}^3
H_{j}H_{k})- G_{5X}\dot{\phi}^3
H_{j}H_{k}(H_{j}+H_{k})\,,\,\,{\cal G} \equiv
2G_4-2G_{4X}\dot{\phi}^2+G_{5\phi}\dot{\phi}^2 \,.&&
\end{eqnarray}
A dot means a derivative with respect to $t$; $H_i=\dot { a}_i/{
a}_i$  -- the Hubble parameters;
$H=\dfrac{1}{3}\sum\limits_{i=1}^3 H_i\equiv \dot{ a}/{ a}$  --
the average Hubble parameter with ${ a}=({ a}_1{ a}_2{
a}_3)^{1/3}$. The triples of indices $\{i,j,k\}$ take values
$\{1,2,3\}$, $\{2,3,1\}$, or $\{3,1,2\}$.

The equation for the scalar field $\phi(t)$ can be represented as
\begin{eqnarray}\label{scalar}
\frac{1}{\rm a^3}\frac{d}{dt}({\rm a^3} {\cal J})={\cal P}\,,
\end{eqnarray}
with $$ {\cal J} =\dot{\phi}\, \Big[ G_{2X}-2G_{3\phi}
+3H\dot\phi(G_{3X} -2G_{4X\phi})+$$
\begin{eqnarray} \label{J}+G^0_0(-2G_{4X}-2\dot\phi^2G_{4XX} +2G_{5\phi}
+G_{5X\phi}\dot\phi^2 )
 +H_1H_2H_3(3G_{5X}\dot{\phi} +G_{5XX}\dot{\phi}^3)\Big]\,,\end{eqnarray}
 $${\cal P} = G_{2\phi} -\dot{\phi}^2(G_{3\phi\phi}
+G_{3X\phi}\ddot \phi) +RG_{4\phi}+2G_{4X\phi}\dot\phi(3\ddot\phi
H-\dot\phi G^0_0)+ $$ \begin{eqnarray}
\label{P}+G^0_0G_{5\phi\phi}\dot{\phi}^2+G_{5X\phi}\dot{\phi}^3H_1H_2H_3\,.
\end{eqnarray}

For convenience, we consider the parametrization of the metric
\begin{equation}
ds^2 =
-dt^2+a^2(t)[e^{2(\beta_{+}+\sqrt{3}\beta_{-})}dx^2+e^{2(\beta_{+}-\sqrt{3}\beta_{-})}dy^2+e^{-4\beta_+(t)}dz^2].
\label{met}
\end{equation}
The functions $e^{\beta_{+}+\sqrt{3}\beta_{-}}$\,,
$e^{\beta_{+}-\sqrt{3}\beta_{-}}$ and $e^{-2\beta_{+}}$ are the
deviation from isotropy, and $a(t)$ is the isotropic part. The
Hubble parameters  in the direction of $x$, $y$ and $z$ are given
by
\begin{equation}
H_1=H+\dot{\beta}_{+}+\sqrt{3}\dot{\beta}_{-}\,,\,
H_2=H+\dot{\beta}_{+}-\sqrt{3}\dot{\beta}_{-}\,, \,
H_3=H-2\dot{\beta}_{+}\,. \label{H}
\end{equation}
We choose the subclass of HG:
\begin{equation}\label{Gi} G_{2}=G_2(X)\,,\, G_3=G_3(X)\,,\, G_4=\frac{1}{16\pi}\,,\,  G_5=\dfrac{\eta\phi}{2}\,.\end{equation}
Equations (\ref{00}), (\ref{ii}) and (\ref{scalar}) will change as
follows
 \begin{eqnarray}3\big(H^2-\sigma^2\big)\left(\frac{1}{8\pi}+\frac{3\eta\dot{\phi}^2}{2}\right) =-G_2 +\dot{\phi}^2G_{2X}+ 3G_{3X}H\dot{\phi}^3
\label{tad00}\,,
\end{eqnarray}
\begin{eqnarray}\big(2\dot{H}+3H^2 + 3\sigma^2\big)\left(\frac{1}{8\pi}+\frac{\eta\dot{\phi}^2}{2}\right)+\eta H\cdot\frac{d(\dot{\phi}^2)}{dt} =
-G_2 +G_{3X}\dot{\phi}^2\ddot{\phi} ,  \label{dh} &&
\\
\label{bb}
\dot{\beta}_{\pm}\left(\frac{1}{8\pi}+\frac{\eta\dot{\phi}^2}{2}\right)
 = \frac{C_{\pm}}{{ a}^3}, \, \, \sigma^2 \equiv \dot{\beta}^2_{+} + \dot{\beta}^2_{-}\,,&&
\end{eqnarray}
\begin{eqnarray}\dot{\phi}\, \Big[ G_{2X}+3HG_{3X} \dot\phi-3\eta (H^2-\sigma^2)\Big]
  \label{scalartad00}
 =\frac{C_0}{{ a}^3}, \end{eqnarray} where  $C_0$, $C_+$ and $C_-$ are the integration constants. Constants $C_{\pm}$ correspond to the
anisotropic charges; $C_0$ is the scalar charge.  The system
(\ref{tad00}), (\ref{dh}), (\ref{bb}) and (\ref{scalartad00})
contains three independent equations and five functions ($H(a)$,
$X(a)$, $\sigma^2$, $G_2$, $G_3$). There is freedom to choose two
functions out of five.  For example, we can set the law of the
the Universe development $H(a)$  and  the canonical kinetic term
$X(a)$. Then we find  the functions $\dot{\beta}_{\pm}$, $G_2(X)$
and $G_3(X)$. If all charges are simultaneously $C_0=C_{\pm}=0$,
then the system (\ref{tad00}), (\ref{dh}), (\ref{bb}),
(\ref{scalartad00}) can only have a stationary solution $H,\, X=
\text{const}$, $\dot{\beta}_\pm=0$. Equality $\dot{\beta}_\pm=0$
means that space-time has become the FRW spacetime. In case
$C_0\neq 0$, $C_{\pm}\neq 0$, we will have a rich variety of
cosmological models.

Equations (\ref{tad00}), (\ref{dh}) and (\ref{scalartad00}) have
consequences
\begin{equation}\label{recg2}G_2=-3\left(\frac{1}{8\pi}+\eta X\right)(H^2[a(X)]-\sigma^2[a(X)])+\frac{\varepsilon C_0\sqrt{2X}}{a^3(X)}\,,\end{equation}
\begin{equation}\label{recg23X}G_{3X}=\frac{2\eta H}{\varepsilon\sqrt{2X}}+\frac{2}{\varepsilon\sqrt{2X}}\left(\frac{1}{8\pi}+\eta X\right)
\left(H'_X+\frac{3a'_X\sigma^2}{a(X)H}\right)+\frac{C_0a'_X}{Ha^4(X)}\,,\end{equation}
where $\varepsilon=\pm 1$ defines the sign of the derivative
$\dot\phi=\varepsilon\sqrt{2X}$. It is easy to verify that
$\dot{\phi}\ddot{\phi}=\dot{X}=aH/a'_X$. As can be seen from the
corollary of equation (\ref{bb})
\begin{equation}\label{siqmkin}\dot{\beta}_{\pm}=\frac{C_\pm}{a^3\left(\frac{1}{8\pi}+\eta X(a)\right)}
 \,,\end{equation}
the anisotropic properties of the Universe are directly determined
by the choice of the density $X(a)$.

\section{Reduction to the isotropic case}

To begin with, we consider reconstruction in isotropic space. If
we set $\dot{\beta}_{\pm}$ to zero, then we obtain the isotropic
spacetime:
\begin{equation}
ds^2 = -dt^2+a^2(t)[dx^2+dy^2+dz^2]\,, \label{}
\end{equation}
then the field equations (\ref{tad00}), (\ref{dh}) and
(\ref{scalartad00}) take the form
\begin{eqnarray}3H^2\left(\frac{1}{8\pi}+\frac{3\eta\dot{\phi}^2}{2}\right)
=-G_2 +\dot{\phi}^2G_{2X}+ 3G_{3X}H\dot{\phi}^3 \label{}\,,
\end{eqnarray}
\begin{eqnarray}\big(2\dot{H}+3H^2)\left(\frac{1}{8\pi}+\frac{\eta\dot{\phi}^2}{2}\right)+\eta H\cdot\frac{d(\dot{\phi}^2)}{dt} =
-G_2 +G_{3X}\dot{\phi}^2\ddot{\phi} ,  \label{} &&
\end{eqnarray}
\begin{eqnarray}\dot{\phi}\, \Big[ G_{2X}+3HG_{3X} \dot\phi-3\eta H^2\Big]
  \label{scalartad01}
 =\frac{C_0}{{ a}^3}\,. \end{eqnarray}
The restored functions (\ref{recg2}), (\ref{recg23X}) will take
the form
\begin{equation}\label{recg20}G_2=-3\left(\frac{1}{8\pi}+\eta X\right)H^2[a(X)]+\frac{\varepsilon C_0\sqrt{2X}}{a^3(X)}\,,\end{equation}
\begin{equation}\label{recg23X0}G_{3X}=\frac{2\eta H}{\varepsilon\sqrt{2X}}+\frac{2H'_X}{\varepsilon\sqrt{2X}}\left(\frac{1}{8\pi}+\eta X\right)
+\frac{C_0a'_X}{Ha^4(X)}\,.\end{equation}

When considering the modified theories of gravity, one has to
worry about the presence of ghosts as well as Laplacian
instabilities \cite{DeFelice1,Appleby1}. The scalar perturbations
will be stable if \cite{DeFelice1}:
\begin{equation}c_S^2\equiv\frac{3(2w_1^2w_2H-w^2_2w_4+4w_1w_2\dot w_1-2w_1^2\dot w_2)}{w_1(4w_1w_3+9w_2^2)}\geq 0\,,\label{cS}
\end{equation}
\begin{equation}
Q_{S}\equiv\frac{w_{1}(4w_{1}w_{3}+9w_{2}^{2})}{3w_{2}^{2}}>0\,,
\label{QS}
\end{equation}
where
\begin{eqnarray}
&&w_{1}  \equiv  2\,(G_{{4}}-2\,
XG_{{4X}})-2X\,(G_{{5X}}{\dot{\phi}}H-G_{{5\phi}}) \,,
\label{w1def}\\
&&w_{2}  \equiv  -2\, G_{{3X}}X\dot{\phi}+4\,
G_{{4}}H-16\,{X}^{2}G_{{4{ XX}}}H+4(\dot{\phi}G_{ {4\phi X}}-4H\,
G_{{4X}})X+2\, G_{{4\phi}}\dot{\phi} \nonumber
\\
 &  & \ \ \ \ \ \ \ \,
 +8\,{X}^{2}HG_{{5\phi X}}+2H\, X\,(6G_{{5\phi}}-5\,
G_{{5X}}\dot{\phi}{H})-4G_{{5{
XX}}}{\dot{\phi}}X^{2}{H}^{2}\,,\\
&& w_{3}  \equiv  3\, X(G_{2{X}}+2\, XG_{{
2XX}})+6X(3X\dot{\phi}HG_{{3{ XX}}}-G_{{3\phi X}}
X-G_{{3\phi}}+6\, H\dot{\phi}G_{{3X}})
\nonumber \\
 &  & \ \ \ \ \ \ \ \,
 +18\, H(4\, H{X}^{3}G_{{4{  XXX}}}-HG_{{4}}-5\, X\dot{\phi}G_{{4\phi
X}}-G_{{4\phi}}\dot{ \phi}+7\, HG_{{4X}}X+16\, H{X}^{2}G_{{4{
XX}}}-2\,{X}^{2}\dot{\phi}G_{{4\phi{
XX}}})\nonumber \\
 &  & \ \ \ \ \ \ \ \,    +6{H}^{2}X(2\, H\dot{\phi}G_{{5{
XXX}}}{X}^{2}-6\,{X}^{2}G_{{5\phi{  XX}}}+13XH\dot{\phi}G_{{5{
XX}}}-27G_{{5\phi
X}}X+15\, H\dot{\phi}G_{{5X}}-18G_{{5\phi}})\,,\\
&&w_{4}  \equiv  2G_{4}-2XG_{5\phi}-2XG_{5X}\ddot{\phi}~.
\end{eqnarray}
The speed square $c_S^2>0$ excludes the Laplacian instabilities.
The condition (\ref{QS}) guarantees  the absence of ghosts.
Similarly, for tensor perturbations we have \cite{DeFelice1}:
\begin{equation}Q_T\equiv\frac{w_1}{4}>0\,,\, c^2_{T}\equiv\frac{w_4}{w_1}\geq 0\,.\label{csT}
\end{equation}
Taking into account (\ref{recg20}), (\ref{recg23X0}) and
$G_5=\eta\phi/2$, we rewrite functions $Q_T$, $c^2_{T}$, $c^2_S$
and $Q_S$ in the form
\begin{equation} Q_T=\frac{1}{4}\left(\frac{1}{8\pi}+ \eta X(a)\right)\,,\, c^2_{T}=\frac{1-8\pi\eta \cdot X(a)}{1+8\pi\eta \cdot X(a)}\,,\end{equation}

\begin{equation}Q_S=\frac{3\left(\frac{1}{8\pi}+ \eta X\right)}{w_{2}^{2}}\left[S\left(\frac{1}{\pi}+16\eta X\right)
\left(\frac{X^{5/2}H'_a}{a^4(X'_a)^2H}-\frac{X^{3/2}}{2a^4X'_a}\right)+\frac{4S^2X^3}{a^8(X'_a)^2H^2}\right]\,,\end{equation}

\begin{equation}c^2_S=\frac{1}{Q_S}\left[-\omega_4+2H\left(\frac{\omega_1^2a}{w_2}\right)'_a\right]\,,
\, S=\varepsilon C_0\sqrt{2}\,,\end{equation} where
$w_1=1/(8\pi)+\eta X$, $w_4=1/(8\pi)-\eta X$ and
\begin{equation}w_2=2\left(\frac{1}{8\pi}+ \eta X\right)\left(H-\frac{2H'_aX}{X'_a}\right)-\frac{2SX^{3/2}}{a^4X'_aH}\,.\end{equation}
The Hubble parameter $H(a)$ is set from cosmological
considerations. The choice of the function $X(a)$ makes it
possible to satisfy conditions (\ref{cS}), (\ref{QS}) and
(\ref{csT}) in the region of applicability of the model $H(a)$.

\subsection{Post-inflationary transition to the radiation-dominated
phase}

Now we will show reconstruction example of the gravitating scalar
field theory for a post-inflationary transition to the
radiation-dominated phase. Further, the theory will be tested for
the absence of pathologies.

We define the Hubble parameter $H(a)$ as follows
\begin{equation}\label{hpredp} H^2=\frac{8\pi\rho}{3}\,,\,\rho=\frac{2\rho_d}{a^4+1},\, \rho_d=\text{const}\,.\end{equation}
The effective energy density $\rho$ corresponds to the equation of
state
\begin{equation}\label{ewjjd}p=\frac{\rho}{3}-\mu\rho^2\,,\,
\mu>0\end{equation} describes the transition from the inflationary
era to the radiation era in the the early Universe \cite{Pierre}.
The linear term  $\rho/3$ describes the radiation. The nonlinear
part  $-\mu\rho^2$ may be due to Bose-Einstein condensates with
self-interaction.  In case (\ref{ewjjd}), the deceleration
parameter (DP) has the form
\begin{equation}\label{InflRaddp}q(\rho)=\frac{d}{dt}\left(\frac{1}{H}\right)-1=1-\frac{3\mu\rho}{2}=1-\frac{\rho}{\rho_d}\,.\end{equation}
The DP changes sign at point $\rho_d$\,:
\begin{equation}  q(\rho_d)=0\,,\, \rho_d=\frac{2}{3\mu}\,.\end{equation}
The Universe is accelerating when $\rho>\rho_d$ and decelerating
when $\rho\leq\rho_d$. The point $a=a_d=1$ corresponds to the end
of inflation.  The density $\rho$ is a bounded function:
\begin{equation}2\rho_d\leftarrow\rho\rightarrow 0\,,\, \,\text{as}\,\, 1\gg a \gg 1\,.\end{equation}
The Planck density $\rho_p$   is approximately equal to
$\rho_p\approx 2\rho_d$. With this in mind, we write
\begin{equation}\rho=\frac{\rho_p}{a^4+1}\,.\end{equation}
In the early time ($a\ll 1$), the model has the quasi-de Sitter
behavior: \begin{equation}H^2 \rightarrow
\frac{8\pi\rho_{p}}{3}=\text{const}\,.\end{equation} At late time
($a\gg 1$) the Universe enters in the radiation era:
\begin{equation}H^2\propto
\frac{1}{a^4}\,.\end{equation} We define the input function
$X(a)$:
\begin{equation}\label{kin0}X=\frac{m}{8\pi \eta a^6}\,,\, m=\text{const}\,.\end{equation}
Looking ahead, we note that the choice of $X(a)$ will determine
the behavior of the anisotropy in the Bianchi I model.

Assumption (\ref{kin0}) allows us to obtain from (\ref{recg20}),
(\ref{recg23X0}) the functions $G_2$ and $G_{3X}$:
\begin{equation}G_2=S\left(\frac{8\pi\eta}{m}\right)^{1/2}X-\frac{\rho_p\cdot X^{2/3}(1+8\pi \eta X)}{\left(\frac{8\pi\eta}{m}\right)^{2/3}+X^{2/3}}\,,\end{equation}
$$G_{3X}=\frac{1}{4\varepsilon }\sqrt{\frac{3}{\pi\rho_{p}}}\cdot X^{-5/6}\left[\left(\frac{8\pi\eta}{m}\right)^{2/3}+X^{2/3}\right]^{1/2}\times$$
\begin{equation}\times\left\{-\frac{S}{6}\left(\frac{8\pi\eta}{m}\right)^{1/2}+
\frac{16\pi\rho_p}{3}\left[\frac{\eta
X^{2/3}}{\left(\frac{8\pi\eta}{m}\right)^{2/3}+X^{2/3}}+\left(\frac{8\pi\eta}{m}\right)^{2/3}\frac{1+8\pi
\eta X}{24\pi
X^{1/3}\left(\left(\frac{8\pi\eta}{m}\right)^{2/3}+X^{2/3}\right)^2}\right]\right\}\,.\end{equation}
During  the radiation era ($a\gg1$) we have a limit: $$G_2\simeq
-\rho_p\left(\frac{m}{8\pi \eta}\right)^{2/3}\cdot X^{2/3}\,,$$
\begin{equation}\label{g3Rad} G_3\simeq -\varepsilon\sqrt{\frac{\rho_p}{3\pi}}\cdot \left(\frac{m}{8\pi \eta}\right)^{1/3}\cdot X^{-1/6}+const\,.\end{equation}
The quasi-de Sitter behavior ($a\ll 1$):
$$G_2\simeq \left[-8\pi\eta \rho_p+S\left(\frac{8\pi\eta}{m}\right)^{1/2}\right]\cdot X\,,$$
\begin{equation}\label{g3Sitter}G_3\simeq \frac{\varepsilon}{2} \sqrt{\frac{3}{\pi\rho_p}}\cdot
\left[\frac{16\pi\eta
\rho_p}{3}-\frac{S}{6}\left(\frac{8\pi\eta}{m}\right)^{1/2}\right]\cdot
X^{1/2}+const\,.\end{equation}

\begin{figure}[h]
\includegraphics[width=6cm]{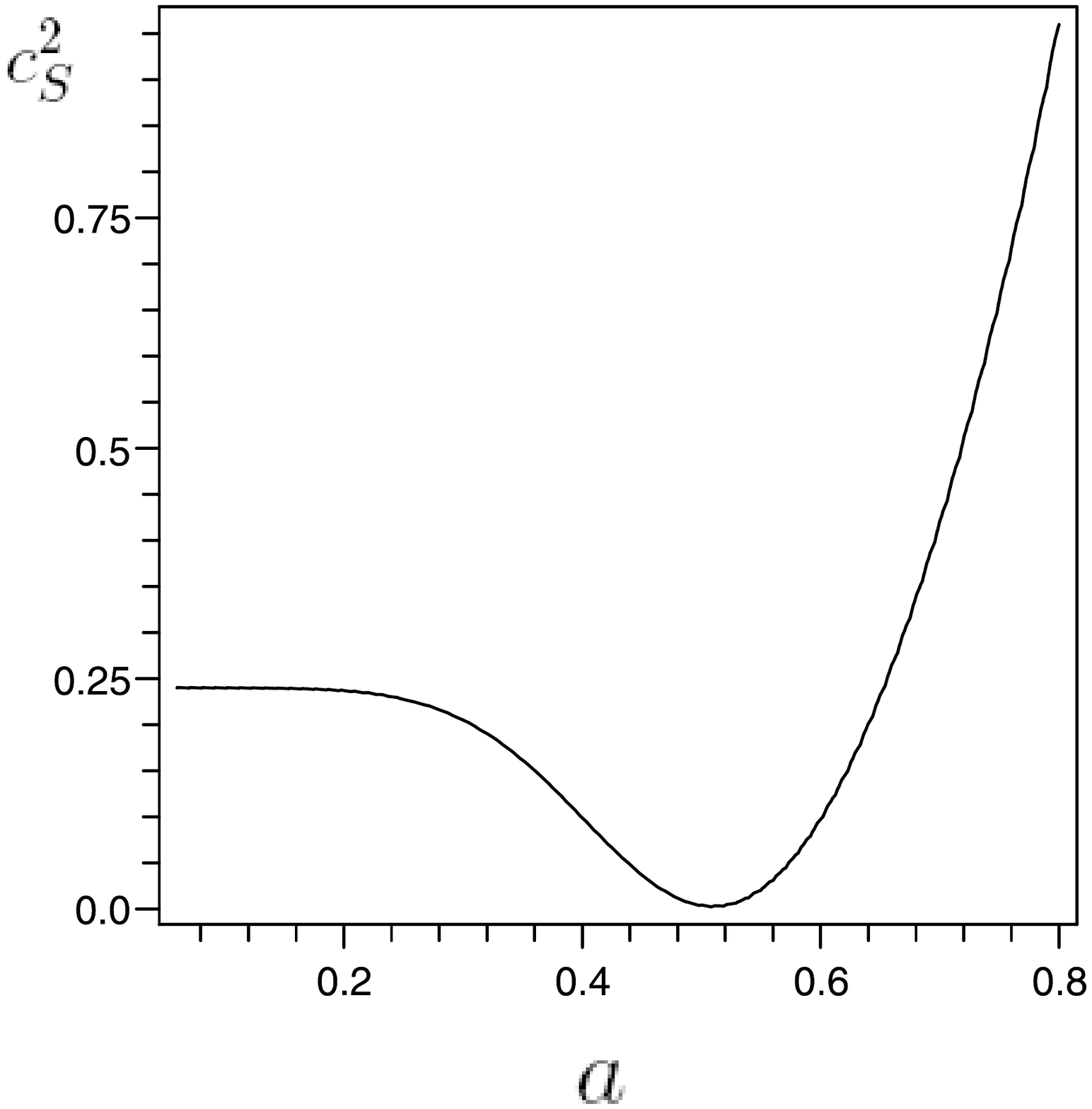}
\caption{Profile of the sound speeds squared $c_S^2$ for $m=0.01$,
$B=1$. The sound speed squared is positive. \label{cs}}
\end{figure}

\begin{figure}[h]
\includegraphics[width=6cm]{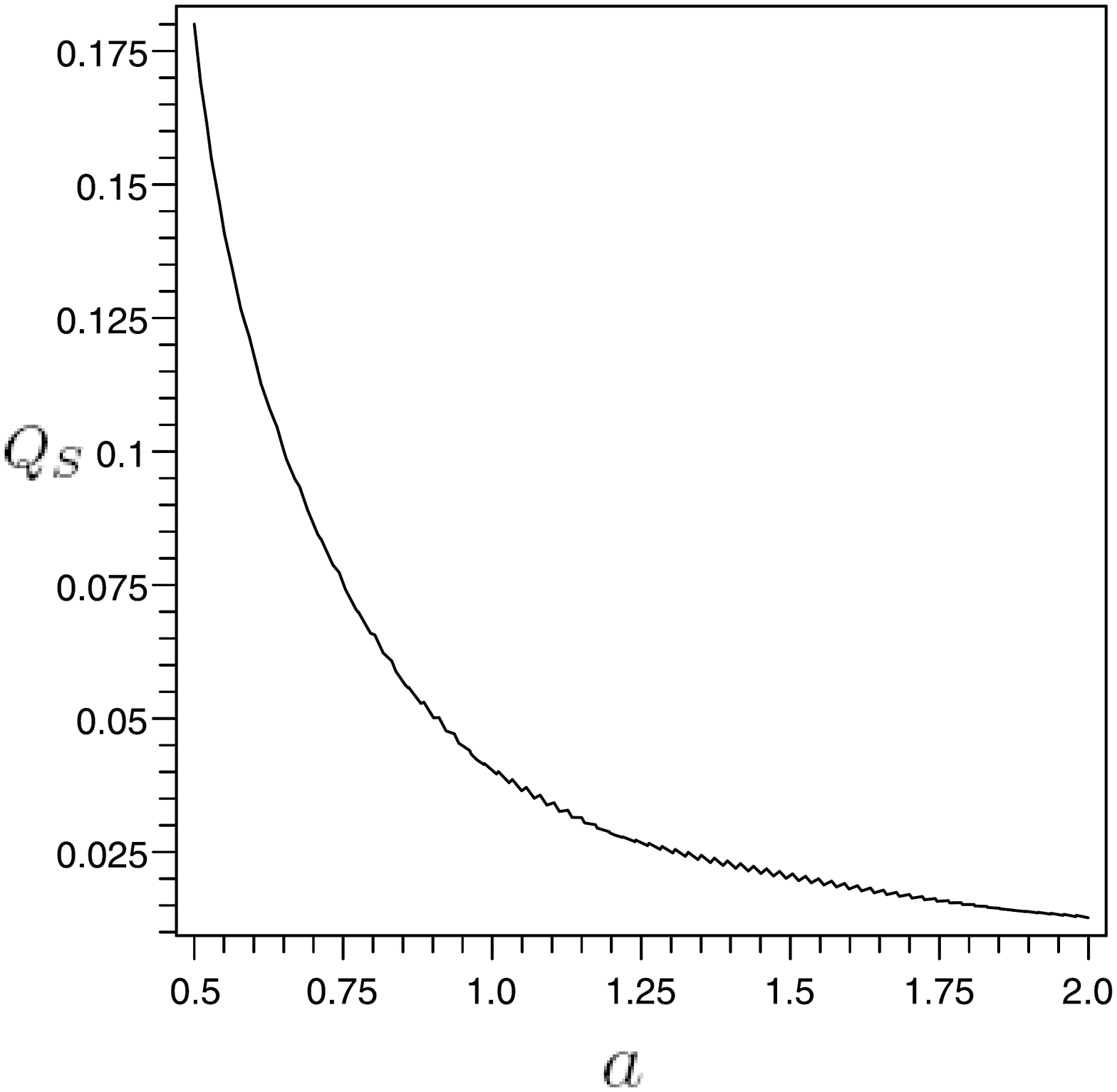}
\caption{Profile of the function $Q_S$ for $m=0.01$, $B=1$. The
function $Q_S$ is positive. \label{qs}}
\end{figure}

For the model under consideration, we rewrite functions $Q_T$,
$c^2_{T}$,  $Q_S$ and $c^2_S$ in the form:
\begin{equation}Q_T=\frac{a^6+m}{32\pi a^6}\,, \, c^2_T=\frac{a^6-m}{a^6+m}\,,\end{equation}
\begin{equation}Q_S=\frac{3B(a^6+m)(1+a^4)^2[2(a^6+2m)(a^4+3)+B(a^4+1)^2]}{8\pi a^6[2(a^6+m)(a^4+3)+B(a^4+1)^2]^2}\,,\end{equation}
\begin{equation}c^2_S=\frac{1}{Q_S}\left[-\omega_4+2H\left(\frac{\omega_1^2a}{w_2}\right)'_a\right]\,,\end{equation}
where
\begin{equation}\omega_1=\frac{a^6+m}{8\pi a^6}\,,\, \omega_4=\frac{a^6-m}{8\pi a^6}\,,\end{equation}
\begin{equation}\omega_2=\frac{A[2(a^6+m)(a^4+3)+B(a^4+1)^2]}{24\pi a^6(1+a^4)^{3/2}}\,, \, A\equiv\sqrt{\frac{8\pi \rho_p}{3}}\,, \,
B\equiv \frac{S}{A^2}\left(\frac{8\pi
m}{\eta}\right)^{1/2}\,.\end{equation} Conditions (\ref{csT}) for
tensor perturbations are satisfied when $a^6>|m|$. In the case
$m>0$, the inequality $Q_T> 0$ holds for any $a>0$. However,
$c^2_T$ has a negative limit, $c^2_T\rightarrow -1$, $a\rightarrow
0$. The function $Q_S$, $c^2_S$ have the following approximations:
\begin{equation}
Q_S\propto\frac{3Bm(12m+B)}{8\pi (6m+B)^2a^6}\,,\, a\rightarrow 0
\,; \,\,Q_S\propto \frac{3B}{8\pi a^2}\,,\, \,a\rightarrow
+\infty\,;\end{equation}
\begin{equation}
c^2_S\rightarrow\frac{(B-24m)(6m+B)}{3B(12m+B)}\,,\, a\rightarrow
0 \,; \,\, c^2_S \propto \frac{8a^2}{3B}\,,\, \,a\rightarrow
+\infty\,.\end{equation} The profiles of functions $c^2_S$, $Q_S$
are shown in Fig. \ref{cs} and Fig. \ref{qs}. Conditions
(\ref{cS}), (\ref{QS}) for the scalar perturbations are satisfied.

\section{Bianchi I model}\label{sec2}

\begin{figure}[h]
\includegraphics[width=6cm]{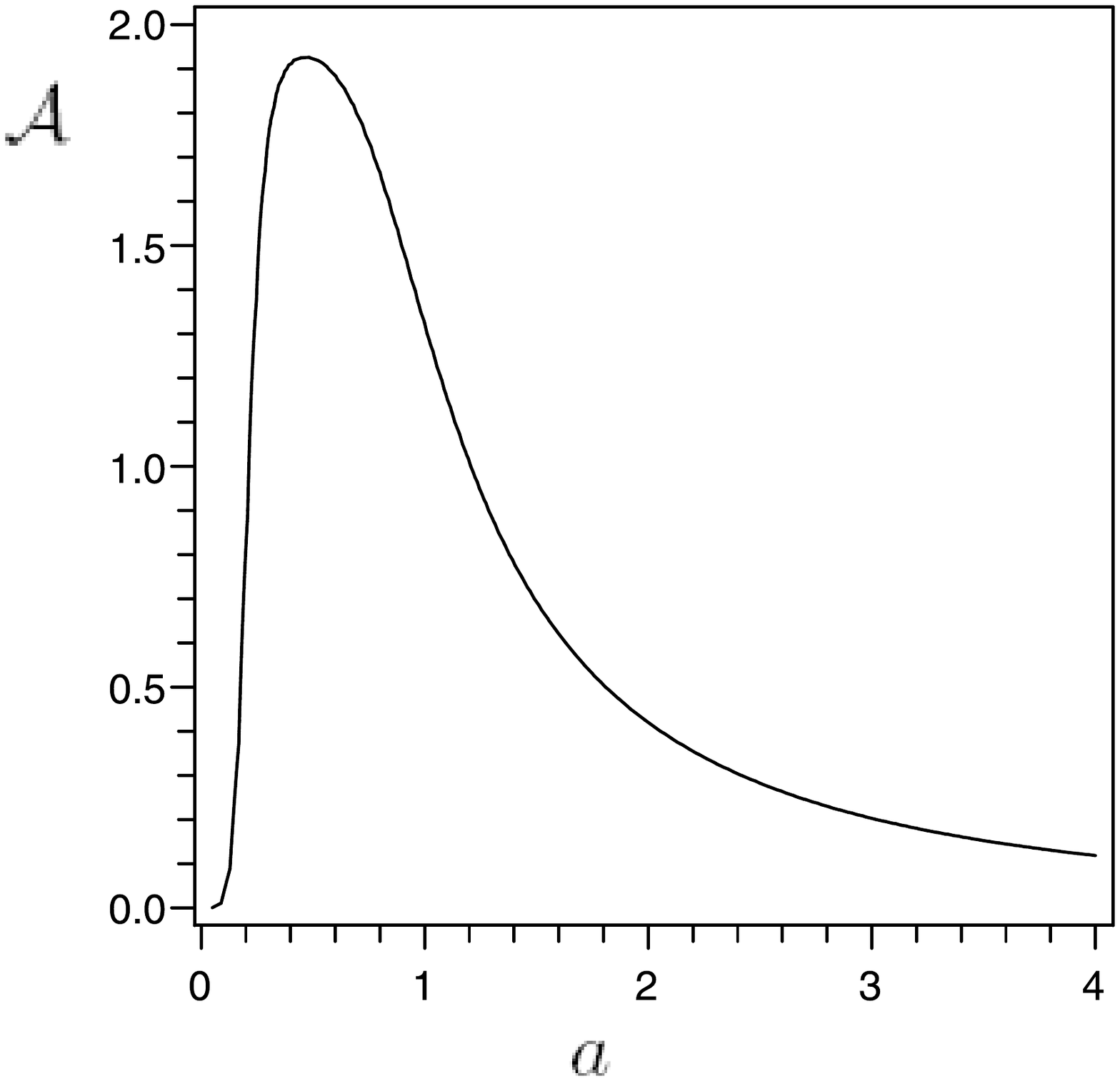}
\caption{Profile of the parameter $\mathcal{A}$ for $m=0.01$,
$F=1$. \label{aniz}}
\end{figure}

\begin{figure}[h]
\includegraphics[width=6cm]{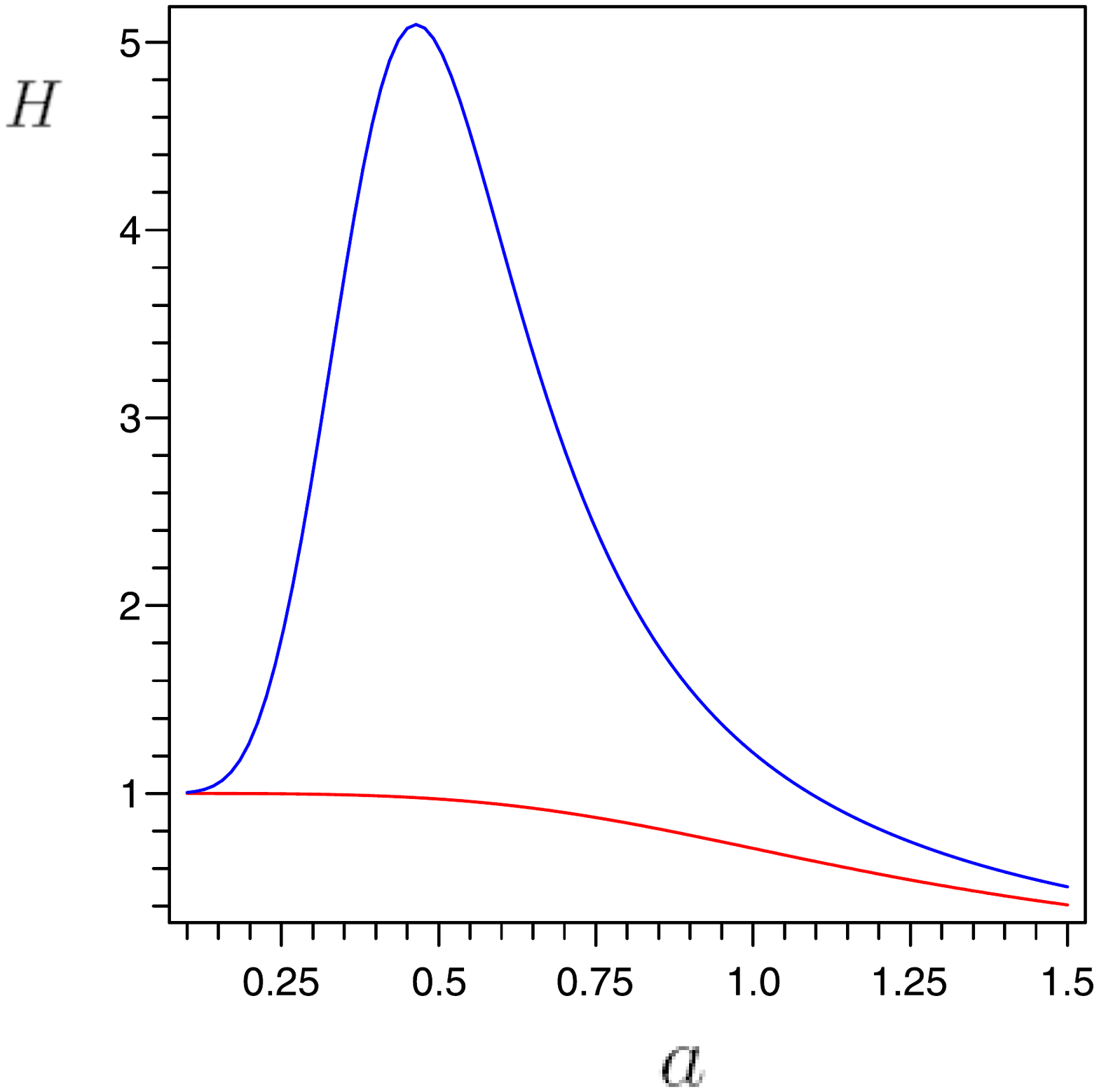}
\caption{Profiles of the parameter
$\sqrt{\frac{3}{8\pi\rho_{p}}}\cdot H$ for isotropic model (red)
and anisotropic model (blue). \label{hhan}}
\end{figure}

\begin{figure}[h]
\includegraphics[width=10cm]{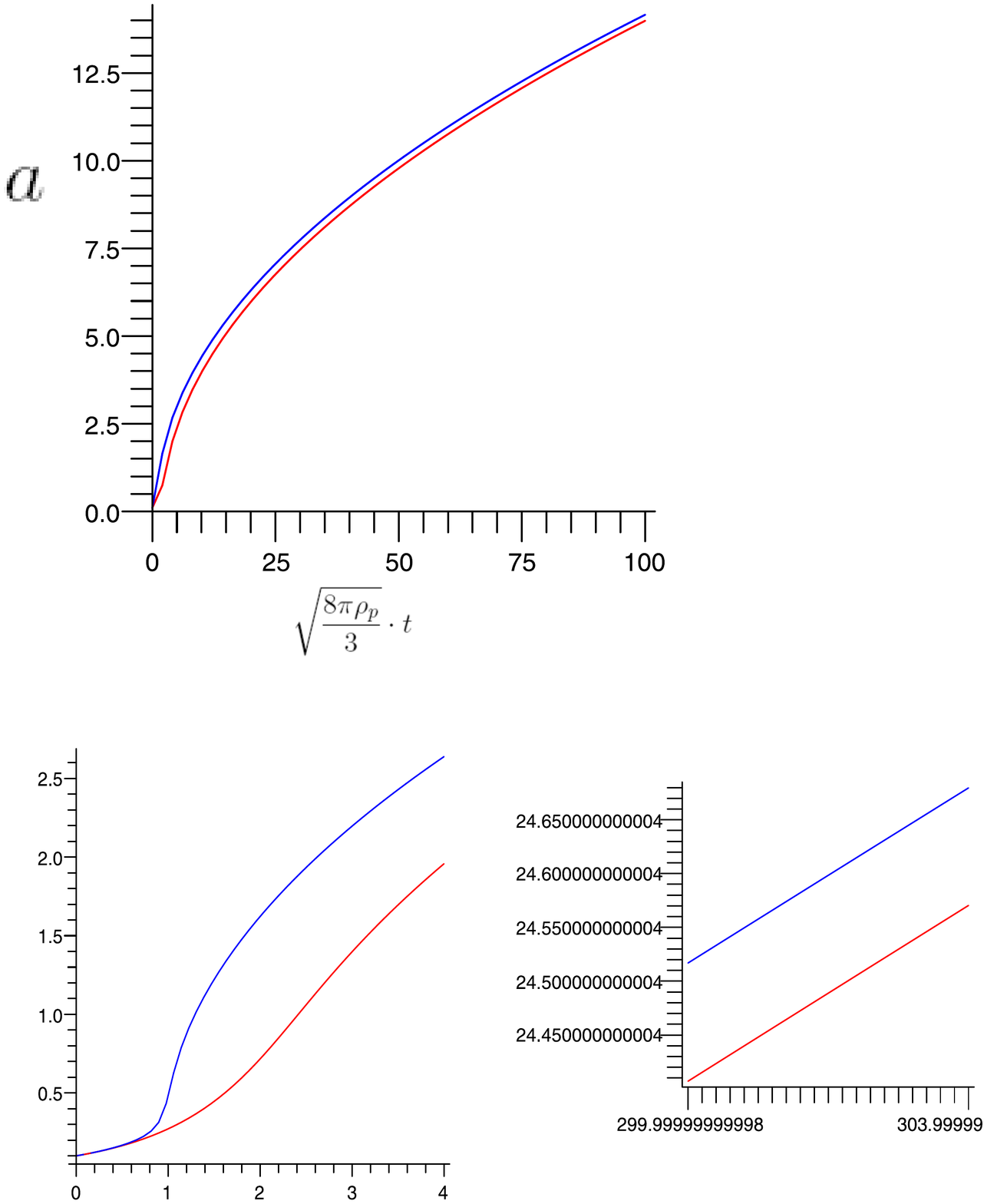}
\caption{Scalar factor profiles for isotropic model (red) and
anisotropic model (blue) with the initial values $a(0)=0.1$.
\label{factall}}
\end{figure}

Let us consider an anisotropic generalization of the model
(\ref{hpredp}), (\ref{kin0}). The assumption (\ref{kin0}) for the
function $X(a)$ gives a corollary to equation (\ref{siqmkin}):
\begin{equation}\dot{\beta}_{\pm}=\frac{8\pi C_\pm \cdot a^3}{a^6+m}\,,\,
\sigma^2 =\frac{(8\pi)^2\sigma^2_0a^6}{(a^6+m)^2}\,,\end{equation}
where $\sigma^2_0\equiv C_-^2+C_+^2$.  In the general theory of
relativity, spatial anisotropies produce in the Einstein equations
terms proportional to $1/a^6$, which become dominant when one goes
backwards in time.  Here, the shear scalar $\sigma^2$ is a
non-monotonic function with the properties
\begin{equation}\sigma^2\rightarrow0\, \, \text{as} \,\, 0\leftarrow a\rightarrow\infty\,.\end{equation}
The shear scalar tend to zero for $a \rightarrow 0$ and their
contribution to the total energy balance is $\propto a^6$ instead
of $\propto1/a^6$.  This is the non-standard behavior at small
$a$.  The shear scalar is suppressed at late time as well by the
factor $1/a^6$. Taking into account anisotropy, we generalize the
assumption (\ref{hpredp}):
\begin{equation}\label{anizhpredp} H^2=\sigma^2+\frac{8\pi\rho}{3}\,,\end{equation}
where
\begin{equation}\rho=\frac{\rho_p}{a^4+1}\,.\end{equation}
The mean anisotropy parameter $\mathcal{A}$ is defined as
$$\mathcal{A}=\frac{1}{3}\sum_{i=1}^{3}\left(\frac{H_i-H}{H}\right)^2=\frac{2\sigma^2}{H^2}
=$$\begin{equation}=2\left[1+\frac{F(a^6+m)^2}{a^6(a^4+1)}\right]^{-1},\,
F\equiv\frac{\rho_p}{24\pi\sigma^2_0}\,.\label{aniz1}\end{equation}
This parameter describes the isotropization process  of the
Universe. The function $\mathcal{A}$ has a limit:
$\mathcal{A}\rightarrow0\, \, \text{as} \,\, 0\leftarrow
a\rightarrow\infty$. Therefore, the anisotropy effect is totally
negligible at early and late times. The behavior of the parameter
$\mathcal{A}$ is shown in Fig. \ref{aniz}. As we can see, the
selected function $X(a)$ provides isotropization in later times.
Since the anisotropy is suppressed at early and late times, the
behavior of the model (\ref{anizhpredp}) is similar to that of the
model (\ref{hpredp}) at these times:
\begin{equation}H^2\approx
\frac{8\pi\rho}{3}\rightarrow
\frac{8\pi\rho_{p}}{3}=\text{const}\,, \, a\ll 1\,.\end{equation}
\begin{equation}H^2\approx
\frac{8\pi\rho}{3}\propto \frac{1}{a^4}\,, a\gg 1\,.\end{equation}
There is a difference for models in intermediate times. The
quasi-de Sitter stage and the radiation era are separated by the
anisotropic phase. The profile  Fig.\ref{hhan} compares the
behavior of the Hubble parameter for models (\ref{hpredp}) and
(\ref{anizhpredp}). The scale factors $a$ of the two models are
shown in Fig.\ref{factall}.

Assumption (\ref{kin0}) allows us to obtain from (\ref{recg2}),
(\ref{recg23X}) the functions $G_2$ and $G_{3X}$:
\begin{equation}G_2=S\left(\frac{8\pi\eta}{m}\right)^{1/2}X-\frac{\rho_p\cdot X^{2/3}(1+8\pi \eta X)}{\left(\frac{8\pi\eta}{m}\right)^{2/3}+X^{2/3}}\,,\end{equation}
$$G_{3X}=\frac{1}{\varepsilon \sqrt{2X}}\left(\frac{(8\pi)^3\sigma^2_0\eta X}{m(1+8\pi \eta X)^2}+
\frac{8\pi\rho_{p}}{3}\cdot
\frac{X^{2/3}}{\left(\frac{8\pi\eta}{m}\right)^{2/3}+X^{2/3}}\right)^{-1/2}\times$$
\begin{equation}\times\left\{-\frac{S}{6}\left(\frac{8\pi\eta}{m}\right)^{1/2}+
\frac{16\pi\rho_p}{3}\left[\frac{\eta
X^{2/3}}{\left(\frac{8\pi\eta}{m}\right)^{2/3}+X^{2/3}}+\left(\frac{8\pi\eta}{m}\right)^{2/3}\frac{1+8\pi
\eta X}{24\pi
X^{1/3}\left(\left(\frac{8\pi\eta}{m}\right)^{2/3}+X^{2/3}\right)^2}\right]\right\}\,.\end{equation}
Here, approximations (\ref{g3Rad}), (\ref{g3Sitter}) are also
valid. Stability conditions (\ref{cS}), (\ref{QS}), (\ref{csT})
have been derived for the homogeneous and isotropic backgrounds.
They can be used also in the Bianchi I model case at early and
late times, since the anisotropies are then damped.


\section{Conclusion}

We have proposed a way to reconstruct the HG theory  on the Hubble
evolution and on the canonical kinetic term $X$ in the Bianchi I
spacetime model. The experience of many researchers makes it
possible  to set  a scenario for the evolution of the Universe
{\it a priori}. The freedom of choice $X(a)$ is used to regulate
the stability of the model and the anisotropic properties of the
Universe.

We have reconstructed the scalar theory, which gives the model of
post-inflationary transition to the radiation-dominated phase in
the Bianchi I spacetime. As a basis for the reconstruction, we
took a fluid with the nonlinear equation of state. At  the
Universe stages with suppressed anisotropy, this equation has the
form $p=\rho/3-\mu\rho^2$. The term $-\mu\rho^2$ may be due to
Bose-Einstein condensates with repulsive ($\mu < 0$) or attractive
($\mu > 0$) self-interaction. The kinetic dependence $X(a)\sim
a^{-6}$ gives the following properties for the model. It is
important that in the process of expansion, the Universe is
isotropized. Interestingly, the anisotropy is also suppressed in
the early times of the Universe. At intermediate times, there is a
finite burst of anisotropy with further relaxation. The isotropic
analogue of the model does not contain ghosts and Laplacian
instabilities for scalar perturbations. For the tensor
perturbations, on the interval $(0, a_1)$, the square of the sound
speed $c^2_T$  is negative. For the anisotropic model, these
judgments are valid at stages of the Universe with suppressed
anisotropy.

\acknowledgments This work is supported by the Russian Foundation
for Basic Research (Grant No. 19-52-15008).

\end{document}